\documentclass[aps,prl,preprint,superscriptaddress,showpacs,preprintnumbers,reprint,amssymb,linenumbers]{revtex4-1}

\usepackage{graphicx}
\usepackage[latin1]{inputenc}
\usepackage{braket}

\usepackage{units}
\usepackage{wrapfig}
\usepackage{graphicx}
\usepackage{amsmath}
\usepackage{booktabs}
\usepackage{rotating}
\usepackage{color}
\usepackage{amsfonts}
\usepackage{amssymb}
\usepackage{moreverb}
\usepackage{url}
\usepackage{epsf}
\usepackage{multirow}
\usepackage{subfigure}
\usepackage{morefloats}
\usepackage{threeparttable}
\usepackage{cancel}
\usepackage{bm}
\usepackage[version=3]{mhchem}
 \usepackage{subfigure}
\usepackage{booktabs}
\usepackage{array}
\usepackage{sidecap}
\usepackage{units}

\begin{document}

\title{Boron aggregation in the ground states of boron-carbon fullerenes}
\author{Stephan Mohr}
\affiliation{Institut f\"ur Physik, Universit\"at Basel, Klingelbergstr. 82, 4056 Basel, Switzerland}
\author{Pascal Pochet}
\affiliation{Laboratoire de simulation atomistique (L\_Sim), SP2M, UMR-E CEA / UJF-Grenoble 1, INAC, Grenoble, F-38054, France}
\author{Maximilian Amsler}
\affiliation{Institut f\"ur Physik, Universit\"at Basel, Klingelbergstr. 82, 4056 Basel, Switzerland}
\author{Bastian Schaefer}
\affiliation{Institut f\"ur Physik, Universit\"at Basel, Klingelbergstr. 82, 4056 Basel, Switzerland}
\author{Ali Sadeghi}
\affiliation{Institut f\"ur Physik, Universit\"at Basel, Klingelbergstr. 82, 4056 Basel, Switzerland}
\author{Luigi Genovese}
\affiliation{Laboratoire de simulation atomistique (L\_Sim), SP2M, UMR-E CEA / UJF-Grenoble 1, INAC, Grenoble, F-38054, France}
\author{Stefan Goedecker}
\affiliation{Institut f\"ur Physik, Universit\"at Basel, Klingelbergstr. 82, 4056 Basel, Switzerland}

\date{\today}

\begin{abstract}
We present novel structural motifs for boron-carbon nano-cages of the stochiometries \ce{B12C48} and \ce{B12C50}, based on first principle calculations. These configurations are distinct from those proposed so far by the fact that the boron atoms are not isolated and distributed over the entire surface of the cages, but rather aggregate at one location to form a patch. Our putative ground state of \ce{B12C48} is  \unit[1.8]{eV} lower in energy than the previously proposed ground state and violates all the suggested empirical rules for constructing low energy fullerenes. The \ce{B12C50} configuration is energetically even more favorable than \ce{B12C48}, showing that structures derived from the \ce{C60} Buckminsterfullerene are not necessarily magic sizes for heterofullerenes. 
\end{abstract}


\maketitle

Since its discovery by Kroto \textit{et al.}\ in 1985 \cite{Kroto1985C60}, the \ce{C60} fullerene has found a wide range of applications as a building block in the field of nano-science. For instance it is possible to directly form solids out of it \cite{kratschmer1990solid} or to dope it by adding substitutional or endohedral atoms, e.g.\ in the context of hydrogen storage \cite{zhao2005hydrogen}.

For future applications it would be advantageous to have more such basic building blocks. 
One possibility is to modify the original carbon fullerene by substitutional doping. 
A very popular choice for the dopant atoms are boron and nitrogen since they are neighbors to carbon in the periodic table.
Various boron-carbon heterofullerenes have been observed experimentally~\cite{guo1991doping,dunk2013formation}. The existence of cross-linked \ce{N12C48} fullerenes could explain experimental measurements of thin solid films~\cite{hultman2001cross}.
The case of boron is of particular importance as it is the p-type counterpart of the n-type nitrogen doping in fullerene and graphene used to tune their electronic or catalytic properties~\cite{wang2012review,xie2003tailorable}.
 



In order to determine the energetically most favorable structure, it is in principle necessary to perform a numerically costly 
unbiased global structure search, which became possible only very recently~\cite{pickard2011ab,glass2006uspex,wang2012calypso,Goedecker2004}. 

Previous work on heterofullerene cages was therefore based on biased searches that constrained the exploration of the energy landscape to certain 
structural motifs. 
For the stochiometries \ce{B_nC_{60-n}} this constraint typically consisted in starting the search from the perfect \ce{C60} fullerene and substituting $n$ carbon atoms by boron. 
Garg \textit{et al.}\ \cite{Garg2011substitutional} extensively investigated the geometries \ce{B_nC_{60-n}} for $n=1-12$. They concluded that the boron is arranged such that a pentagon ring does not contain more than one and a hexagon not more than two boron atoms (at non-adjacent sites). Putting more boron atoms in a ring increases the bond lengths and decreases the stability. A study by Viani and Santos \cite{Viani2006comparative} on various smaller fullerenes again confirmed that boron atoms are most preferably situated at opposite sites in a hexagon, thereby increasing the bond lengths in their neighborhood. For the heterofullerene \ce{B12C48}, which will be one of the two stochiometries investigated in this paper,  Manaa \textit{et al.}\ \cite{RiadManaa2003} performed a detailed study; they claimed that the best structure was the same that was previously found for \ce{N12C48} \cite{Manaa2002}, thereby again confirming the previous results stating that the boron atoms should be spread over the entire carbon cage and isolated. We will refer to this class of structures as ``diluted''.

We reinvestigated the boron-carbon heterofullerenes by performing for the first time a systematic and unbiased search for low 
energy minima for the stochiometries \ce{B12C48} and \ce{B12C50}. 

The Minima Hopping Method (MHM)~\cite{Goedecker2004} was used for our unbiased search of new configurations. 
All calculations were done at the level of Density Functional Theory, using the BigDFT package \cite{Genovese2008}. This code uses Daubechies wavelets \cite{daubechies1992ten} as basis set and HGH pseudopotentials \cite{Hartwigsen1998} to simulate the core electrons. The exchange correlation part was described by the PBE functional \cite{Perdew1996}, which has shown to give highly reliable energy differences between different structural motifs in boron \cite{hsing2009quantum} and is therefore used~\cite{marques2012libxc} in this work. 
Convergence parameters in BigDFT were set such that total energy differences were converged up to \unit[$10^{-5}$]{Ha}  
and all configurations were relaxed until the maximal force component on any atom reached the noise level of the calculation, which was of the order of \unit[1]{meV/\r{A}}.

In order to speed up our exhaustive configurational search we started both from diluted and aggregated configurations with a wide structural diversity, based on various \ce{C_n} backbones (with $n$=60, 58 or 48). For aggregated configurations we considered both filled hexagons and filled pentagons as they were demonstrated to be the building blocks~\cite{pochet2011low,boulanger2013selecting} in the \ce{B80} boron fullerene~\cite{GonzalezSzwacki2007}; furthermore we have inserted compact boron icosahedrons in opened \ce{C60} fullerenes. In addition we generated structures of high symmetry for the stoichiometry \ce{C48} and added 12 boron atoms at locations where it seemed appropriate by intuition. Some short MHM runs starting from the energetically most promising configurations allowed us to see trends leading to an energy lowering. From these ones a second set was derived by applying additional modifications (e.g.\ exchanging  boron and  carbon atoms) in order to speed up the exploration in a second step of MHM. Finally we systematically exchanged boron and carbon atoms up to second-nearest neighbors for the most favorable structures emerging from this process.
In this way we could generate more than thousand configurations containing a large variety of structural motifs.


Among our \ce{B12C48} structures obtained in this way we found several being considerably lower in energy than the three most favorable ones identified by Manaa \textit{et al.}\ \cite{RiadManaa2003}. Whereas the new configurations agree with them in the overall shape exhibiting a cage-like structure, they differ substantially by the fact that the boron is not distributed over the entire cluster, but aggregated in a patch, thereby separating the surface of the compound in a boron-rich and a boron-poor part. This is in strong contrast to the widely accepted belief that
the boron atoms should be isolated \cite{Shakib2011}, i.e.\ be always separated by one or several carbon atoms.
We will refer to our novel structural motif as ``patched''.

\begin{figure}
 \centering
 \includegraphics[width=0.49\textwidth]{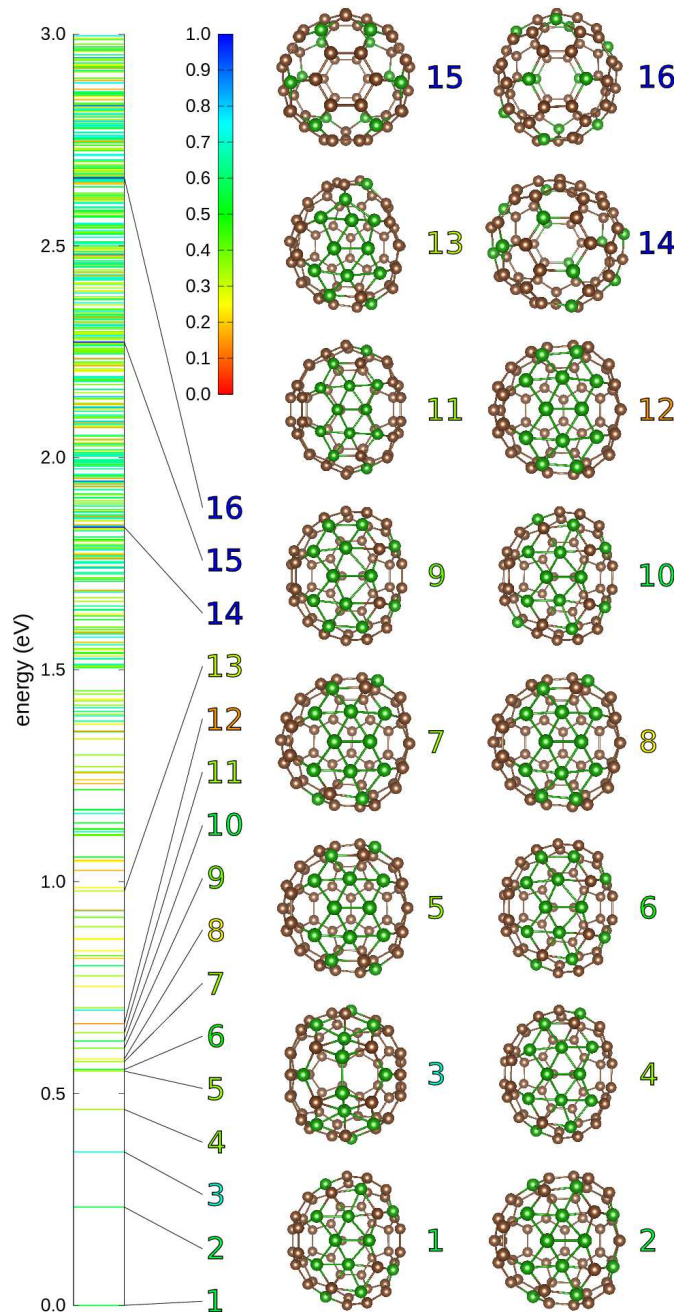}
 \caption{Plot of the 12 energetically most favorable structures, the most favorable one containing a filled heptagon and the three lowest diluted configurations for \ce{B12C48}, together with the low energy part of the spectrum. The coloring scheme is explained in the text.}
 \label{fig:B12C48_spectrum-with-figures_paper}
\end{figure}
The low energy part of the spectrum that we have explored is shown in Fig.\ \ref{fig:B12C48_spectrum-with-figures_paper}, together with the structures of the 12 lowest isomers that we have found, the lowest structure exhibiting a filled pentagon and the three most favorable diluted structures. The latter ones turned out to be identical to the three lowest structures identified by Manaa \textit{et al.} In total we found 143 new structures that are lower in energy than the putative ground state proposed so far. We made however no effort to find all the structures in this energy interval and so more than 143 structures are expected to exist. The energy levels are colored on a scale from 0 to 1 which describes the relative amount of carbon atoms being first neighbors to boron. Thus a value of 0 (red) means that the boron atoms are only surrounded by boron -- which can obviously not happen --, whereas a value of 1 (blue) means that the boron atoms are only surrounded by carbon. Consequently the coloring of the patched structures tends towards red values, while the coloring of the diluted ones tends towards blue values.
In the lowest structures the boron atoms form a patch with frayed boundaries, which results in astonishing configurations where carbon atoms have four boron atoms as first neighbors.
Furthermore it is surprising that the lowest structure exhibits a heptagon, which is usually less favorable than the penta- and hexagons, and twice two adjacent pentagons, which is in general very disadvantageous \cite{kroto1987the}. However these pentagons do not only consist of carbon, but there are some substitutional boron atoms contained within them.


Tab.\ \ref{tab:B12C48_data} gives some more information about the structures shown in Fig.\ \ref{fig:B12C48_spectrum-with-figures_paper}. The first column shows the energy separation $\Delta E$ of the configurations with respect to the lowest structure. We can see that our new ground state is \unit[1.8]{eV} lower in energy than the putative ground state proposed so far. 
\begin{table}
\centering
\begin{tabular}{l r m{6pt} r m{6pt} r m{6pt} l m{6pt} r}
\hline
        & \multicolumn{1}{c}{$\Delta E$} && \multicolumn{1}{c}{gap}  && \multicolumn{1}{c}{$\Delta H$} && \multicolumn{1}{c}{PG} && \multicolumn{1}{c}{RMSD}    \\
        & \multicolumn{1}{c}{(\unit{eV})}&& \multicolumn{1}{c}{(\unit{eV})} && \multicolumn{1}{c}{(\unit{eV})}  &&               && \multicolumn{1}{c}{(\unit{bohr})} \\
          \cline{2-2}                       \cline{4-4}                 \cline{6-6}                       \cline{8-8}               \cline{10-10}
  1st   & 0.000                          && 0.457                    && 0.340                        && $\text{C}_\text{s}$                     &&  0.000                           \\
  2nd   & 0.232                          && 0.477                    && 0.344                        && $\text{C}_\text{s}$                     &&  1.003                           \\
  3rd   & 0.362                          && 0.646                    && 0.346                        && $\text{C}_\text{s}$                     &&  1.717                           \\
  4th   & 0.462                          && 0.645                    && 0.348                        && $\text{C}_\text{1}$                     &&  1.092                           \\
  5th   & 0.553                          && 0.295                    && 0.350                        && $\text{C}_\text{s}$                     &&  1.589                           \\
  6th   & 0.557                          && 0.236                    && 0.350                        && $\text{C}_\text{1}$                     &&  0.558                           \\
  7th   & 0.575                          && 0.247                    && 0.350                        && $\text{C}_\text{1}$                     &&  1.486                           \\
  8th   & 0.582                          && 0.452                    && 0.350                        && $\text{C}_\text{1}$                     &&  1.579                           \\
  9th   & 0.607                          && 0.604                    && 0.350                        && $\text{C}_\text{s}$                     &&  0.783                           \\
  10th  & 0.624                          && 0.378                    && 0.351                        && $\text{C}_\text{1}$                     &&  0.533                           \\
  11th  & 0.644                          && 0.616                    && 0.351                        && $\text{C}_\text{s}$                     &&  1.344                           \\
  12th  & 0.665                          && 0.644                    && 0.351                        && $\text{C}_\text{s}$                     &&  1.490                           \\
  27th  & 0.978                          && 0.695                    && 0.357                        && $\text{C}_\text{1}$                     &&  1.678                           \\
  144th & 1.836                          && 0.494                    && 0.371                        && $\text{S}_\text{6}$                     &&  3.465                           \\
  295th & 2.273                          && 0.316                    && 0.378                        && $\text{D}_\text{3d}$                    &&  3.501                           \\
  453th & 2.661                          && 0.302                    && 0.385                        && $\text{S}_\text{6}$                     &&  3.442                           \\
\hline
\end{tabular}
 \caption{Properties of the structures shown in Fig.\ \ref{fig:B12C48_spectrum-with-figures_paper}.}
\label{tab:B12C48_data}
\end{table}
In the second column we present the HOMO-LUMO gaps of the structures. The values range from \unit[0.2]{eV} to \unit[0.7]{eV}, thus being rather small, and do not exhibit any special pattern. In particular there is no notable difference between the class of the patched and the diluted structures.
The next column shows the formation energies per atom $\Delta H$ with respect to the bulk conformations of boron and carbon ($\alpha$-boron and cubic diamond, respectively), which is defined by $\Delta H = (E - n_B E_B^0 - n_C E_C^0)/(n_B+n_C)$ with $E$ being the energy of the compound and $E_X^0$ and $n_X$ being the energy per atom of the reference configurations and the number of atoms, respectively. In the next column we give the point groups of the structures. Whereas the diluted structures are rather symmetric (point groups $\text{S}_\text{6}$ and $\text{D}_\text{3d}$), the new structures are of much lower symmetry ($\text{C}_\text{s}$ and $\text{C}_\text{1}$).
Finally we present the RMSD~\cite{1302.2322} of the structures with respect to the new putative ground state. It is obvious that there is a clear separation between the patched structures and the diluted ones. Within the patched ones, however, there is no relation between the energy and the RMSD.

\begin{figure}
 \begin{center}
 \subfigure[\ce{B12C48}\label{fig:B12C48_energy_vs_RMSD}]{\includegraphics[width=.22\textwidth]{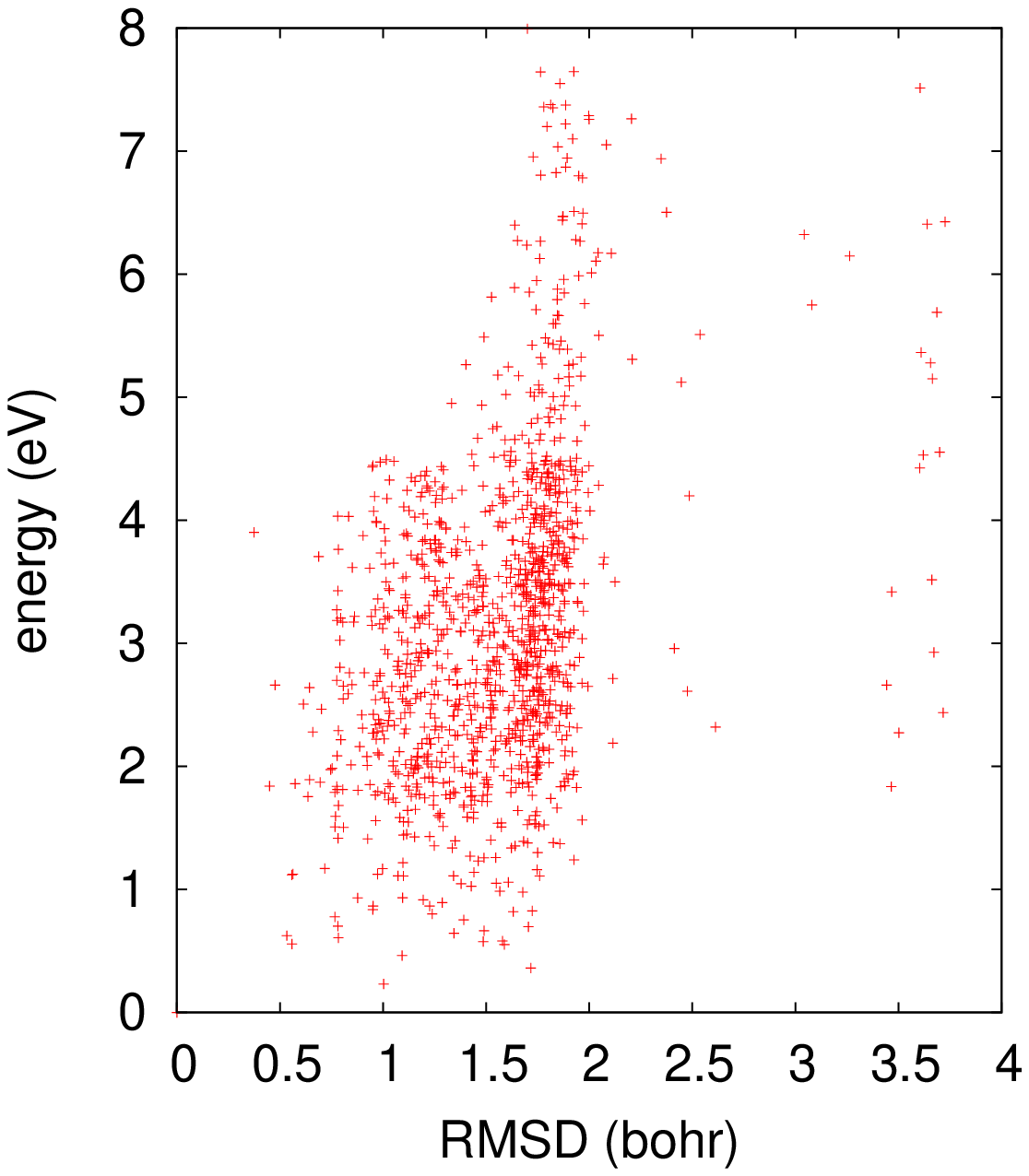}}
 \hspace{12pt}
 \subfigure[\ce{B12C50}\label{fig:B12C50_energy_vs_RMSD}]{\includegraphics[width=.22\textwidth]{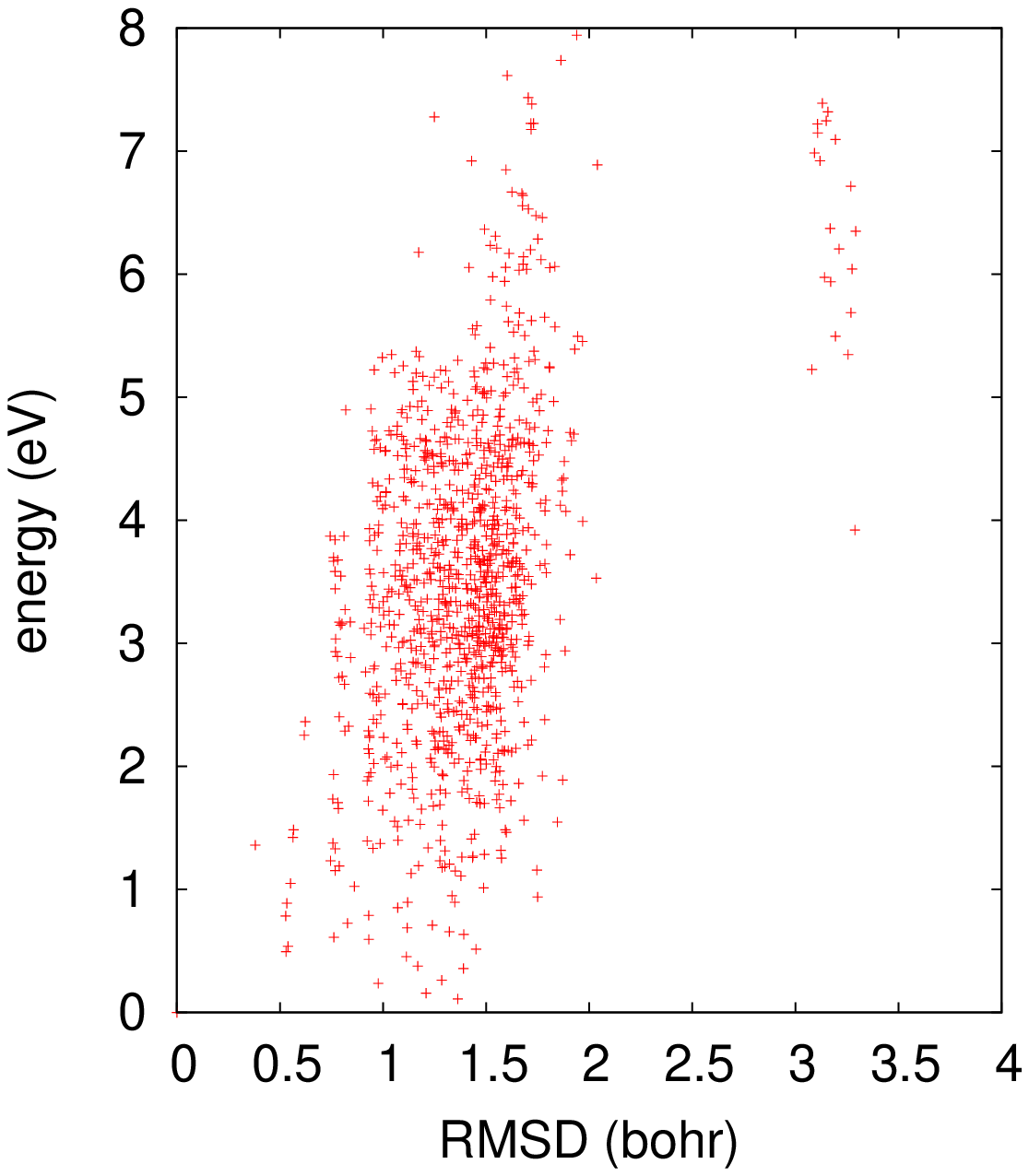}}
 \caption{The energy difference versus the RMSD, both with respect to the putative ground states.}
  \label{fig:energy_vs_RMSD}
 \end{center}
\end{figure}
A more complete overview of the energies and the corresponding RMSD values is shown in Fig.\ \ref{fig:B12C48_energy_vs_RMSD}. Here we plot for all structures up to \unit[8]{eV} the energy difference and the RMSD with respect to the putative ground state. As can be seen there is a broad range with structures whose energies are completely uncorrelated to the value of the RMSD. However this range is sharply bounded at a value of about \unit[2]{bohr}, separating it from energetically higher configurations.

By filling the heptagon in the putative ground state of \ce{B12C48} by two carbon atoms -- thus modifying it into a pentagon and a hexagon -- the two adjacent pentagons which are present twice on both sides of the heptagon are turned into a pentagon and a hexagon each. Since this structure consists only of pentagons and hexagons and furthermore respects the isolated pentagon rule, it is a 
promising  candidate for the global minimum of the stoichiometry \ce{B12C50}.

In order to confirm this assumption we again performed an extended and unbiased search for the ground state. We again used several approaches to generate many different input structures: We added two atoms to the lowest isomers of \ce{B12C48} as described; we replaced ten carbon atoms by boron in a \ce{C60} and added two additional interstitial boron atoms at appropriate positions; and we replaced 12 carbon atoms by boron starting from a \ce{C62} fullerene. For this last approach we took two examples that were first described by Ayuela \textit{et al.}\ \cite{Ayuela1996} and Qian \textit{et al.}\ \cite{Qian2003} and turned out to be the most stable \ce{C62} isomers in a study by Cui \textit{et al.}\ \cite{Cui2007}. 
As for the case of \ce{B12C48} we again continued by either manually rearranging the atoms or by using the configurations as input for subsequent MHM runs. 

\begin{figure}
 \includegraphics[width=0.49\textwidth]{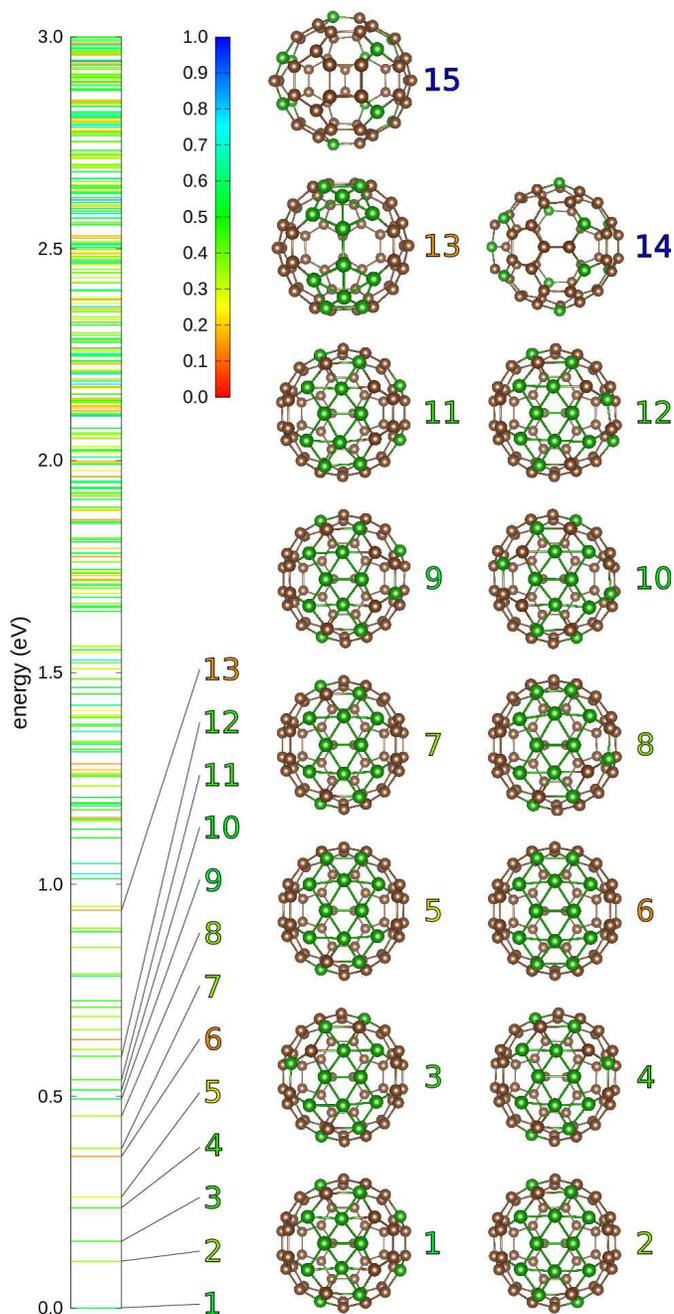}
 \caption{The low energy part of the spectrum for \ce{B12C50}, together with figures of the 12 energetically lowest structures, the most favorable one containing a filled pentagon and the two most favorable diluted configurations.}
 \label{fig:B12C50_spectrum-with-figures_paper}
\end{figure}
It turned out that the structure that we constructed by hand from the ground state of \ce{B12C48} is energetically the most favorable one. 
In addition, at least 673 patched structures  are lower in energy than the most favorable diluted configuration 
which is \unit[3.9]{eV} higher in energy than our ground state. 
The low energy part of the spectrum is shown in Fig.\ \ref{fig:B12C50_spectrum-with-figures_paper}, 
together with figures of the 12 structures being lowest in energy, the most favorable one containing a filled pentagon and the two most favorable diluted configurations. The energy levels are coded in colors in analogy to Fig.~\ref{fig:B12C48_spectrum-with-figures_paper}. As for the case of \ce{B12C48} it is clear that all low lying minima correspond to the same structural motif where the boron is aggregated in a patch. 

In Tab.\ \ref{tab:B12C50_data} we present some more information about the structures shown in Fig.\ \ref{fig:B12C50_spectrum-with-figures_paper}.
\begin{table}
\centering
\begin{tabular}{l r m{6pt} r m{6pt} r m{6pt} l m{6pt} r}
\hline
       & \multicolumn{1}{c}{$\Delta E$} && \multicolumn{1}{c}{gap}  && \multicolumn{1}{c}{$\Delta H$} && \multicolumn{1}{c}{PG} && \multicolumn{1}{c}{RMSD}   \\
       & \multicolumn{1}{c}{(\unit{eV})}       && \multicolumn{1}{c}{(\unit{eV})} && \multicolumn{1}{c}{(\unit{eV})}       &&                        && \multicolumn{1}{c}{(\unit{bohr})} \\
         \cline{2-2}                       \cline{4-4}                 \cline{6-6}                       \cline{8-8}               \cline{10-10}
 1st   & 0.000                          && 0.534                    && 0.322                        && $\text{C}_\text{s}$                     && 0.000                      \\
 2nd   & 0.110                          && 0.416                    && 0.324                        && $\text{C}_\text{2}$                     && 1.363                      \\
 3rd   & 0.158                          && 0.434                    && 0.325                        && $\text{C}_\text{1}$                     && 1.209                      \\
 4th   & 0.237                          && 0.283                    && 0.326                        && $\text{C}_\text{1}$                     && 0.976                      \\
 5th   & 0.262                          && 0.338                    && 0.327                        && $\text{C}_\text{1}$                     && 1.285                      \\
 6th   & 0.358                          && 0.308                    && 0.328                        && $\text{C}_\text{2v}$                    && 1.390                      \\
 7th   & 0.377                          && 0.154                    && 0.328                        && $\text{C}_\text{s}$                     && 1.168                      \\
 8th   & 0.453                          && 0.215                    && 0.330                        && $\text{C}_\text{1}$                     && 1.113                      \\
 9th   & 0.494                          && 0.249                    && 0.330                        && $\text{C}_\text{1}$                     && 0.530                      \\
 10th  & 0.515                          && 0.380                    && 0.331                        && $\text{C}_\text{1}$                     && 1.451                      \\
 11th  & 0.539                          && 0.227                    && 0.331                        && $\text{C}_\text{1}$                     && 0.539                      \\
 12th  & 0.595                          && 0.363                    && 0.332                        && $\text{C}_\text{1}$                     && 0.931                      \\
 25th  & 0.939                          && 0.296                    && 0.337                        && $\text{C}_\text{2v}$                    && 1.750                      \\
 673th & 3.921                          && 0.204                    && 0.386                        && $\text{C}_\text{s}$                     && 3.289                      \\
 965th & 5.227                          && 0.458                    && 0.407                        && $\text{C}_\text{s}$                     && 3.080                      \\
\hline
\end{tabular}
\caption{The same data as in Tab. \ref{tab:B12C48_data}, this time for the structures shown in Fig.\ \ref{fig:B12C50_spectrum-with-figures_paper}.}
\label{tab:B12C50_data}
\end{table}
The HOMO-LUMO gaps are -- except for the few lowest structures -- in general  smaller than the ones for \ce{B12C48}; again there is no notable difference between the patched and the diluted structures.
The values for the formation energy $\Delta H$ are slightly lower than their 
counterparts for \ce{B12C48} in Tab.\ \ref{tab:B12C48_data}, indicating that it is more likely to encounter experimentally 
the stoichiometry \ce{B12C50} than \ce{B12C48}.
The point groups exhibit a wider variation compared to the case of \ce{B12C48}; the energetically most favorable structures have the point groups $\text{C}_\text{1}$, $\text{C}_\text{2}$, $\text{C}_\text{s}$ and $\text{C}_\text{2v}$. On the other hand the diluted structures are of lower symmetry than their counterparts for \ce{B12C48} and exhibit only the point group Cs.
The RMSD behaves similar as for \ce{B12C48}. 
A more complete overview of the energy versus the RMSD -- analogous to the one shown in Fig.\ \ref{fig:B12C48_energy_vs_RMSD} -- is given in Fig.\ \ref{fig:B12C50_energy_vs_RMSD}. 
Both the last column of Tab.~\ref{tab:B12C50_data} and Fig.\ \ref{fig:B12C48_energy_vs_RMSD} show again a clear separation between the patched and the diluted configurations. 

In conclusion, we have shown that the ground state of boron-carbon heterofullerenes is fundamentally different from what was thought previously. It consists of incomplete carbon cages whose openings are filled with boron patches. In a broader context our results show that doping in $\text{sp}^2$-materials is not yet well understood and that no universally valid rules are available to predict which structural motifs are the most stable ones in such doped structures. Our results could also give guidance to synthesis efforts~\cite{de2011energy} for such heterofullerenes. The steep rise of the energy of metastable configurations as a function of the distance from the ground state shown in Fig.~\ref{fig:energy_vs_RMSD} indicates that there is a substantial driving force towards low energy motifs with patches and suggests that a synthesis of patched structures should be possible. The energy gap between the ground state and the lowest metastable state (\unit[0.2]{eV} for \ce{B12C48} and \unit[0.1]{eV} for \ce{B12C50}) is however much smaller than in \ce{C60} (\unit[1.6]{eV}) and reaching the ground state might therefore be difficult. The standard synthesis procedure of substituting~\cite{dunk2013formation} carbon atoms by boron atoms is unlikely to succeed for heterofullerenes containing a larger number of boron atoms which then form patches. Planar boron clusters that are structurally similar to the boron patches found in our ground state can however be synthesized experimentally~\cite{zhai2003an} and might form growth nuclei for such heterofullerenes in a carbon rich spark or vapor chamber.\\

We acknowledge support from the SNF and HP2C. Computing time was provided by the CSCS and GENCI (grant 6194).

\bibliography{citationlist-new}{}
\bibliographystyle{ieeetr}
\end{document}